\documentclass[aps,
		prl,
		reprint,
		twocolumn,
		superscriptaddress,
		shortbibliography,
		nofootinbib,
		floatfix,
		notitlepage
		]{revtex4-2}
\pdfoutput=1

\usepackage[dvipsnames]{xcolor}

\usepackage{graphicx}
\usepackage[colorlinks=true, urlcolor=blue, citecolor=blue, linkcolor=blue]{hyperref}
\usepackage{color}
\usepackage{bm}
\usepackage{newtxtext, newtxmath}
\usepackage{tabularx}
\usepackage{geometry}		
\usepackage{setspace}
\usepackage{cleveref}
\usepackage{braket}
\usepackage{slashed}

\newcommand{\be}{\begin{equation}}
\newcommand{\ee}{\end{equation}}
\newcommand{\bea}{\begin{eqnarray}}
\newcommand{\eea}{\end{eqnarray}}
\newcommand{\bi}{\begin{itemize}}
\newcommand{\ei}{\end{itemize}}

\newcommand{\Tfi}{\tsf{T}_{\tiny\tsf{fi}}}
\newcommand{\Tvac}{\tsf{T}_{\tiny\tsf{vac}}}
\newcommand{\Pvac}{\tsf{P}_{\tiny\tsf{vac}}}
\newcommand{\Wvac}{\tsf{W}_{\tiny\tsf{vac}}}
\newcommand{\Mfi}{\mathfrak{M}_{\tiny\tsf{fi}}}
\newcommand{\vkap}{\varkappa}
\newcommand{\eps}{\varepsilon}
\newcommand{\vphi}{\varphi}

\hypersetup{colorlinks=true,citecolor=blue,urlcolor=magenta}

\newcommand{\mbf}[1]{\mathbf{#1}}
\newcommand{\tbf}[1]{\textbf{#1}}
\newcommand{\trm}[1]{\textrm{#1}}
\newcommand{\tsf}[1]{\textsf{#1}}

\newcommand{\figref}[1]{Fig. \ref{#1}}
\newcommand{\eqnref}[1]{Eq. (\ref{#1})}

\newcommand{\av}[1]{\langle #1 \rangle }

\newcommand{\nn}{\nonumber}

\definecolor{bkcol}{rgb}{0.14,0.42,0.9}
\definecolor{aaaa}{rgb}{0.1, 0.5, 0.1}
\definecolor{bbbb}{rgb}{0.5, 0.3, 0.9}
\definecolor{grayA}{rgb}{0.5, 0.5, 0.5}

\begin{document}

\title{Vacuum muon decay and interaction with laser pulses}

\author{B. King}
\email{b.king@plymouth.ac.uk}
\affiliation{Centre for Mathematical Sciences, University of Plymouth, Plymouth, PL4 8AA, United Kingdom}
\author{D. Liu}
\email{di.liu@plymouth.ac.uk}
\affiliation{Centre for Mathematical Sciences, University of Plymouth, Plymouth, PL4 8AA, United Kingdom}

\date{\today}

\begin{abstract}
Muons decay in vacuum mainly via the leptonic channel to an electron, a muon neutrino and an electron antineutrino. Previous investigations have concluded that muon decay can only be significantly altered in a strong electromagnetic field when the muonic strong-field parameter is of order unity, which is far beyond the reach of lab-based experiments at current and planned facilities. In this letter, an alternative mechanism is presented in which a laser pulse affects the vacuum decay rate of a muon \emph{outside} the pulse. Quantum interference between the muon decaying with or without interacting with the pulse generates fringes in the electron momentum spectra and can increase the muon lifetime by up to a factor 2. The required parameters to observe this effect are available in experiments today. 
\end{abstract}

\maketitle

The highest intensity of electromagnetic fields that can be produced in the lab has been increasing in recent years \cite{Yoon:2021ony} and is set to increase still further with several multi-PW lasers operating or in the process of being commissioned or constructed \cite{danson19}. In anticipation of the extended science reach that these facilities will provide, many suggestions have been made about phenomena that may be studied in strong electromagnetic fields (for reviews, see e.g. \cite{ritus85,Kaminski:2009wwd,DiPiazza:2011tq,Narozhny:2015vsb,Gonoskov:2021hwf,Fedotov:2022ely}). One aspect of these phenomena is \emph{non-perturbativity at small coupling}: the fundamental electromagnetic coupling $\alpha \approx 1/137 \ll 1$ is enhanced by the electromagnetic field intensity to be of order unity or larger. In high-power laser labs, an effective coupling to electrons and positrons, $\xi_{e}$, of order unity, corresponding an all-order interaction between background photons and electrons/positrons, can nowadays be routinely accessed \cite{Sarri:2025qng}. 

Recently it has been demonstrated in experiment how high power lasers can be used to study \emph{electroweak} processes. For example laser-wakefield accelerated electron beams have been collided with high-Z solid targets to produce muons in set-ups that are relatively compact compared to traditional muon sources \cite{Schumaker:2018noe,Zhang:2024axy,Calvin:2025huk}. Muon-antimuon pairs are thereby generated in the Coulomb field of the target nuclei via the (two-step) Bethe-Heitler mechanism of an electron emitting a real photon that decays, or via the (one-step) trident process of direct pair production from an electron \cite{Titov:2009cr}, with some fraction also generated through the decay of charged pions. Charged kaons, along with muons and pions, can even be created by relatively weak lasers, for example in nuclear-induced processes on ultra-dense hydrogen H(0) \cite{holmlid2019decay}.

Noting that the effective coupling $\xi_{\mu}$ of a muon of mass $m_{\mu}$ to the electromagnetic field of a laser pulse is $\xi_{\mu}=\delta\,\xi_{e}$ (with $\delta = m_{e}/m_{\mu}\approx 1/207$) and considering high power lasers coming online can potentially reach $\xi_{e} \sim O(1000)$, we see that measuring small-coupling non-perturbativity in muon-laser interactions with $\xi_{\mu}\sim O(1)$ may soon be within experimental reach. Motivated by these developments, we revisit the question of whether electroweak decays involving electromagnetic charges may be modified by intense laser pulses.

Previous work on electroweak decays in electromagnetic backgrounds has focussed mainly on infinitely-extended fields. In the seminal work by Nikishov and Ritus \cite{Nikishov:1964b} the leptonic decay of pions was calculated and in \cite{ritus-jetp69} also the leptonic decay of muons and neutrino emission of electrons in a constant crossed field were studied in detail. Constant crossed fields are particularly relevant when $\xi \gg 1$ because in that regime, the `formation length' of the process is sufficiently short that it is a good approximation to assume the background is locally constant and crossed \cite{nikishov64,DiPiazza:2017raw,Ilderton:2018nws,Seipt:2020diz}. However, because this parameter regime is not likely to be accessible in experiment in the near future, it is not a regime of immediate interest. General arguments have also been made \cite{BECKER1983131} for why the total probability of a charged particle decay cannot be modified by an intense laser pulse, but the calculation was performed in the quasiclassical limit and the analysis was again concerned with decay in the laser pulse itself. The situation was clearly formulated by Narozhny  and Fedotov \cite{Narozhny:2008zz} that an electromagnetic background can only significantly modify the total probability of an electroweak decay  if the effect is: i) classical, changing the trajectory and hence the time dilation of a decaying particle or ii) quantum, such that the muon strong-field parameter, $\chi_{\mu} = e\hbar\sqrt{-(p\cdot F)^{2}}/m_{\mu}^{3}c^{4}$ where $m_{\mu}$ and $p$ are the muon mass and momentum, $F$ is the field tensor and $e$ is the charge on a positron, must be of order unity. Since $\chi_{\mu} = \delta^{3}\chi_{e}$ and experiment can only currently reach $\chi_{e} \sim O(1)$, this would imply electroweak decays can only be influenced by an intense laser in current and near future experiments by a very small, likely undetectable amount. These arguments were demonstrated by explicit calculation for the case of a monochromatic wave background with $\xi_{\mu} \ll 1$ but arbitrary $\xi_{e}$  by Dicus et al. \cite{Dicus:2008nw,Farzinnia:2009gg}.

In the current letter we consider electroweak decay in a laser pulse of \emph{finite} longitudinal extent. This crucial difference allows for the particle to decay before or after interacting with the laser pulse, thus providing two extra routes to decay that are absent when the laser field is infinitely extended, as in previous treatments. We will find these two decay routes interfere, with fringes appearing in the emitted particle momentum spectra. This `which-way interference' of histories can occur for standard quantum electrodynamic processes in strong-fields \cite{King:2010nka,Dumlu:2011rr,Akkermans:2011yn,Ilderton:2019ceq,Ilderton:2020dhs,Krajewska_2021}; here we will see it for a particle decay process.  Focussing on the example of muon decay, we will find the total vacuum decay rate can be suppressed down to 50\% of its usual value. The controlling parameter originates from the change in the \emph{classical} position of the muon due to having interacted with the laser pulse, but the effect is clearly quantum in nature, arising from the interference of different decay pathways. This mechanism circumvents the restrictions that previous analyses have placed on manipulating electroweak decays with strong electromagnetic fields.

\emph{Outline} -- The most common decay of a muon, $\mu^{-}$ is ${\mu^{-} \to e^{-} + \bar{\nu}_{e} + \nu_{\mu}}$, where $e^{-}$ is an electron,  $\bar{\nu}_{e}$ an electron antineutrino and $\nu_{\mu}$ a muon neutrino. At centre of mass energies much lower than the $W$-boson mass, one can employ Fermi's effective four-fermion interaction (we set $\hbar=c=1$ in the following). The vacuum term can be written:
\bea 
\tsf{T}_{00} = \int \mathfrak{M}_{00}\mbox{e}^{iQ\cdot x}d^{4}x;\quad
\mathfrak{M}_{00}= \frac{G}{\sqrt{2}}\,J_{q,\ell,\sigma}J_{p,k}^{\sigma},
\eea
where $Q=q+\ell+k -p$ is the total momentum change ($p$ is the muon momentum, $q$ is the electron momentum, $k$ is the muon neutrino momentum and $\ell$ is the electron anti-neutrino momentum), ${G \approx (293\,\trm{GeV})^{-2}}$ 
is the Fermi constant \cite{ParticleDataGroup:2024cfk}, and
\bea
J_{q,\ell,\sigma} &=& \frac{\bar{u}_{q}\gamma_{\sigma}(1-\gamma_{5})v_{\ell}}{\sqrt{(2V)^{2}q^{0}\ell^{0}}};~~ 
J_{p,k}^{\sigma} = \frac{\bar{u}_{k}\gamma^{\sigma}(1-\gamma_{5})u_{p}}{\sqrt{(2V)^{2}p^{0}k^{0}}}, \nn
\eea
represent currents, with subscripts on spinors labelling momentum.

\begin{figure}[h!!]
\includegraphics[width=7.5cm]{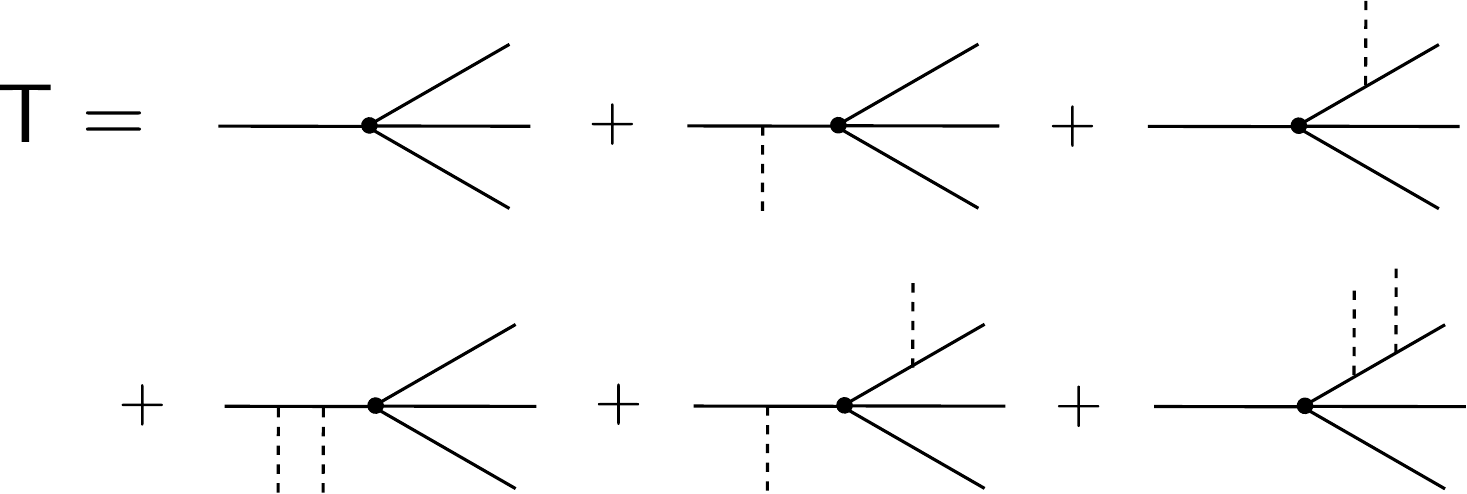}
\caption{First two orders of perturbative expansion in charge-field coupling $\xi$, $\tsf{T}=\tsf{T}_{00}+\tsf{T}_{10}+\tsf{T}_{01}+\tsf{T}_{20}+\tsf{T}_{11}+\tsf{T}_{02}$ respectively.} \label{fig:pert1a}
\end{figure}

\emph{Low intensity background (perturbative case) --} It is instructive to begin with the case that the field intensity is small enough that $\xi \ll 1$, where $\xi$ represents $\xi_{e}$ and $\xi_{\mu}$. Then the  transition matrix $\tsf{T}$ can be expanded in the charge-field interaction $\xi$ as $\tsf{T} = \sum_{ij}\tsf{T}_{ij}$ where $i$ ($j$) refer to the number of interactions between the plane wave background and the muon (electron) (see \figref{fig:pert1a}). The laser background is modelled as a plane wave of finite spatiotemporal extent. The scaled vector potential $a=eA$ (with $e>0$ the charge on a positron) can be written ${a(\phi) = m \xi \eps\,g(\phi)}$ where $m$ is a mass, $\xi$ an intensity parameter, $\eps$ a polarisation vector and $\phi$ the phase where $\phi=\vkap \cdot x$ with $\vkap$ the wavevector. The squared intensity parameter, $\xi_{s}^2$ satisfies  \cite{heinzl10}: $\xi^{2}_{s} = e^{2} \langle p\cdot T \cdot p \rangle_{\phi}/[m_{s}(\vkap\cdot p)]^2$ where $s\in\{e,\mu\}$ refers to the particle species (electron or muon), $p$ is the particle momentum, $T$ is the energy momentum tensor and $\langle \cdot \rangle_{\phi}$ refers to cycle-averaging over the phase, $\phi$. To represent a finite plane wave pulse, we choose $g(\phi)$ to be non-zero only when $0<\phi<\Phi$ so that $\Phi$ denotes the pulse phase duration. (Zero-frequency components can be included in the pulse description, but as we will see, we are interested in channels that do not change the net particle momentum.)
Vacuum decay kinematics are reflected by $\tsf{T}_{00} \propto \delta^{(4)}(Q)$. When external-field photon interactions are added, we find each $\tsf{T}_{ij}$ contains some contribution with the same kinematics as vacuum decay and some with different kinematics, allowing $Q \neq 0$. At the probability level, contributions with the same kinematics will interfere. The vacuum term scales as $\sim VT$ for typical measurement volume $V$ and time $T$, whereas laser-only terms scale as $\sim V \tau$ where $\tau=\Phi/\vkap^{0}$ is the laser pulse duration. Since $\tau \ll T$, only the vacuum terms are retained. Furthermore, the arguments cited in the introduction from previous work also imply direct modification of muon decay \emph{inside} the pulse should be negligible in the perturbative regime.

\begin{figure}[h!!]
\includegraphics[width=4.0cm]{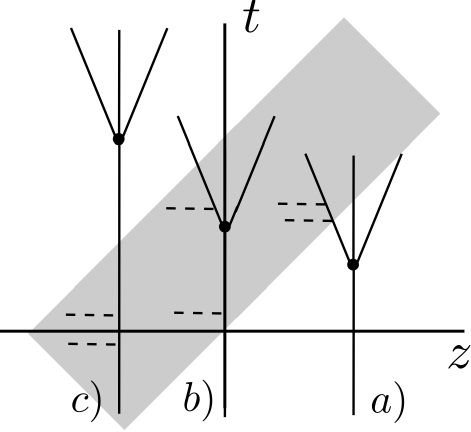}
\caption{Spacetime diagram showing various positions of muon decay, a) before, b) during or c) after laser pulse (grey shaded region), where dashed lines indicate interaction with the pulse.} \label{fig:spacetime1a}
\end{figure}

Adding external-field photons and writing $\tsf{T}=\Tvac(\xi)$ where $\Tvac$ indicates contributions with kinematics identical to those of vacuum decay. The vacuum term of the decay amplitude is:
\[
\Tvac(0) = \tsf{T}_{00}= (2\pi)^{4}\delta^{(4)}(Q)\mathfrak{M}_{00},
\]
and we find up to order $\xi^{2}$ in the charge-field interaction (see Supplementary A \cite{supp} for details):
\bea \Tvac(\xi) = \Tvac(0)\left[1 + \frac{i\xi}{2}\mathcal{I}_{1}+\frac{\xi^{2}}{2}\left( -\frac{1}{2} \mathcal{I}_{1}^{2}  + i\mathcal{I}_{2}\right)\right] \label{eqn:pertEq1}
\eea
with the 
and integrals:
\bea
\mathcal{I}_{1}= \mathcal{I}_{1q}-\mathcal{I}_{1p}; \quad \mathcal{I}_{1r} = m  \frac{r\cdot \eps}{\vkap \cdot r}\int  g(\phi)\,d\phi \equiv \int \mathcal{I}_{1r}'(\vphi) d\vphi \nn \\
\mathcal{I}_{2}= \mathcal{I}_{2p}-\mathcal{I}_{2q}; \quad \mathcal{I}_{2r}=\frac{m^{2}}{2\,\vkap\cdot r}\int g^{2}(\phi)\,d\phi\equiv \int \mathcal{I}_{2r}'(\vphi) d\vphi, \nn \\ \label{eqn:I1I2}
\eea
where $r \in \{p,q\}$. Hence a finite laser pulse can modify the rate of the \emph{vacuum decay channel}. This occurs when the total momentum absorbed from the laser equals the momentum emitted back into the laser. Therefore contributions from diagrams with an odd number of interactions with the laser such as $\tsf{T}_{10}$ and $\tsf{T}_{01}$ should be negligible (unless the laser has an exceptionally wide bandwidth or is heavily chirped). Indeed, these channels contribute to the vacuum decay amplitude with $\mathcal{I}_{1}$ and for a plane wave background with a finite number of cycles and  a symmetric pulse envelope, each of the terms in $\mathcal{I}_{1}$ integrate to zero, i.e. $\mathcal{I}_{1}=0$. Instead, the main modification of the vacuum decay rate is from the ponderomotive term, $I_{2}$. By considering each diagram in \figref{fig:pert1a}, we can infer where the muon decayed, with the situation illustrated in \figref{fig:spacetime1a}. The channel $\tsf{T}_{20}$ must have occurred after the muon entered the laser pulse (possibly already having exited it), $\tsf{T}_{02}$ must have occurred before the electron exited the laser pulse and $\tsf{T}_{11}$ must have occurred when both the muon and electron were in the laser pulse. 
Since $\tsf{T}_{11}$ is proportional to the cross-term in $\mathcal{I}_{1}^{2}$, which is zero we conclude the contribution to vacuum muon decay from processes that can \emph{only} occur \emph{within} the laser pulse, is negligible. Instead, the laser contributes to vacuum decay by interacting with the muon or electron before or after the decay itself. Another way of seeing this is to consider the momentum change $Q(\vphi)$ for a decay in the plane wave pulse, by replacing the muon momentum $p$ and electron momentum $q$ with their plane-wave values. We then find $Q(\vphi)=Q+\Delta Q(\vphi)$ where:
\bea
\Delta Q(\vphi) = \vkap \left[\frac{2 a\cdot q - a\cdot a}{2\vkap\cdot q} - \frac{2 a\cdot p - a\cdot a}{2\vkap\cdot p} \right].
\eea
Therefore $Q(\vphi)=0$ at a finite and discrete set of points inside the pulse, where $p\cdot a(\vphi) = 0$ and $a(\vphi)\cdot a(\vphi)=0$ are fulfilled, compared to outside the pulse, where $\Delta Q=0$ everywhere. Indeed the result in \eqnref{eqn:pertEq1} is independent of the shape of the laser pulse and depends only on the square of the integral of its potential. Further analysis of where the process takes place, is given in Appendices A and B.

\emph{High-intensity background --} In an intense laser pulse, where $\xi \gtrsim 1$, both the muon and electron can become `dressed' in the pulse's electromagnetic field. This charge-laser coupling can be included to all orders of interaction by modelling the pulse as a plane wave and employing Volkov wavefunctions \cite{ritus85}. The free fermion wavefunctions then acquire electromagnetic field-dependent additions, for example:
\[
u_{p}\mbox{e}^{-i p\cdot x} \to \left[1+\frac{\slashed{\vkap}\slashed{a}(\phi)}{2\vkap\cdot p}\right]u_{p}\mbox{e}^{-ip\cdot x + iS_{a,p}(\phi)}
\]
i.e. a spinor-valued prefactor and a nonlinear phase given by:
\bea
S_{a,p}(\phi) = \int_{-\infty}^{\phi}\left[-\xi\mathcal{I}_{1p}'(\vphi) + \xi^{2}\mathcal{I}_{2p}'(\vphi)\right]\,d\vphi \label{eqn:Sa1}
\eea
where $\mathcal{I}_{1p}'(\vphi)$ and $\mathcal{I}_{2p}'(\vphi)$ are defined in \eqnref{eqn:I1I2}.
Motivated by the perturbative analysis, we can make a general argument for how a laser pulse of finite extent can modify the rate  of vacuum decay of electroweak processes involving electromagnetically charged particles. The transition matrix can be written as:
\bea 
\tsf{T}(\xi) = \int \mathfrak{M}[a]\,\mbox{e}^{iQ\cdot x+iS_{a}(\phi)}d^{4}x \label{eqn:Tvac2}
\eea
where we have separated out the combined nonlinear phase from the Volkov wavefunctions with:
\[
S_{a}(\phi) = S_{a,p}(\phi) - S_{a,q}(\phi) =  \int_{-\infty}^{\phi} \xi \mathcal{I}_{1}'(\vphi) + \xi^{2} \mathcal{I}_{2}'(\vphi)~d\vphi.
\]
In a plane wave of finite extent $0<\phi<\Phi$,  $S_{a}(\phi)=0$ before the pulse i.e. when $\phi<0$ and $S_{a}(\phi) = S_{a}(\Phi)$ after the pulse, when $\phi > \Phi$. Therefore after the initial particle has propagated through the pulse the amplitude for decay acquires a constant phase. This contribution interferes with the purely vacuum decay contribution, to modify the total vacuum decay channel. (Although a monochromatic field would also modify the vacuum decay channel, since in that case the field is infinite, there is no interference with the purely vacuum contribution; this interference is the central effect we study here.)

Separating the vacuum contributions into those originating before and after the leading edge of the pulse at $\phi=0$,  we find (see Supplmentary B):
\bea 
\tsf{T}(\xi) &=& (2\pi)^{4}\delta^{(4)}(Q)\mathfrak{M}_{00}F(\xi)  + 2(2\pi)^{3}\delta^{\perp,-}(Q)\left(\cdots\right)~~ \label{eqn:TvacFull1}
\eea
where $\delta^{\perp,-}$ is a delta function in the three conserved momenta in a plane wave background. Using the same arguments as before, we retain only the vacuum channel and assume $\tsf{T}(\xi) =\Tvac(\xi)$, where $\Tvac(\xi) = \Tvac(0) F(\xi)$ with:
\bea 
F(\xi) = \frac{1}{2}\left[1+\mbox{e}^{iS_{a}(\Phi)}\right]; \quad F(0)=1.
\eea
Expanding $F(\xi)$ to quadratic order in $\xi$, we find \eqnref{eqn:TvacFull1} tends to the direct, perturbative result from \eqnref{eqn:pertEq1}. To acquire the probability requires forming $|F(\xi)|^{2}$, and noting that $|F(\xi)|^{2} = \cos^{2}[S_{a}(\Phi)/2]$ $\to 1$ for $\xi \to 0$ but ${|F(\xi)|^{2}\to 1/2}$ when the argument is averaged over, we see already at this stage, the origin of the $50\%$ suppression in the decay rate that can take place. This is similar to the double-slit effect where the observer of the decay products lacks `which-way' information on whether the particle decayed with or without the laser pulse interaction. In this case the which-way information is not the trajectory of which slit is chosen, but rather which history of the particle led to its decay.

\begin{figure}[h!!]
\includegraphics[width=6.5cm]{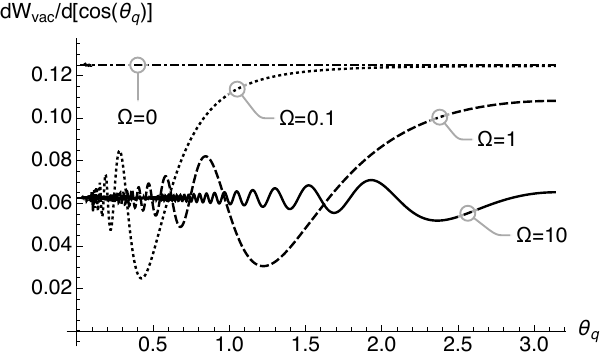}
\caption{Electron polar distribution in the rest frame of the muon $d\Wvac/d[\cos(\theta_{q})]$ for different values of the input parameter $\Omega$.} \label{fig:thetaPlot}
\end{figure}

\begin{figure}[h!!]
\includegraphics[width=6.5cm]{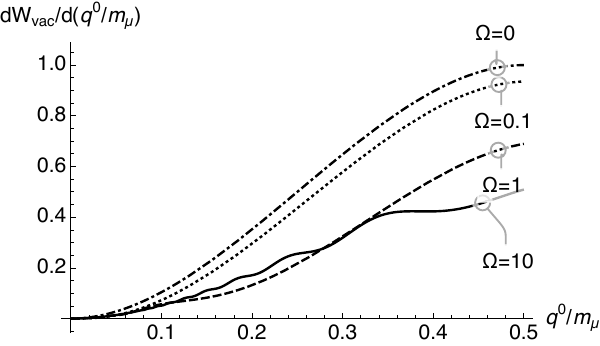}
\caption{Electron energy distribution in the rest frame of the muon $d\Wvac/dq^{0}$ for different values of the input parameter $\Omega$.} \label{fig:EnergyPlot}
\end{figure}

We proceed by calculating the decay rate $\Wvac(\xi)$ (probability per unit time):
\bea 
\Wvac(\xi) = \frac{V^{3}}{T}\int \frac{d^{3}q\,d^{3}\ell\,d^{3}k}{(2\pi)^{9}}~ \big|\Tvac(\xi)\big|^2.
\eea
The derivation proceeds along standard lines (see e.g. \cite{Griffiths:2008}) in the muon rest frame. After the neutrino momenta are integrated over, the electron momentum integral in $q$ is cast in spherical polar co-ordinates with the polar angle, $\theta_{q}$, coinciding with the projection on the laser wavevector, i.e. ${\vkap \cdot q = \vkap^{0}\left(q^{0}-|\mbf{q}|\cos\theta_{q}\right)}$. Performing the trivial integration over the azimuthal angle and setting the electron mass to zero without an appreciable change in the rate (see Supplementary B \cite{supp} for details) leaves:
\bea 
\Wvac(\xi) &=& \frac{G^{2}m_{\mu}^{5}}{48\pi^{3}}\int_{0}^{1/2} dZ \int_{-1}^{1} dX\,Z^{2}(3-4Z)\nn \\
&& \,\qquad\times\left\{1+\cos\left[\Omega\left(1- \frac{1}{Z(1-X)}\right)\right]\right\}
\eea
where $Z=q^{0}/m_{\mu}$, $X=\cos\theta_{q}$ and $\Omega = \xi_{\mu}^{2}\Phi\av{g^{2}}/2\eta_{\mu}$ ($\eta_{\mu}=\vkap \cdot p/m_{\mu}^{2}$ is the muon energy parameter) and ${\av{f}=(1/\Phi)\int_{0}^{\Phi} f(\phi)d\phi}$. The integration in $Z$ and $X$ can be performed analytically but the result is long and not particularly illuminating. In \figref{fig:thetaPlot}, we plot the electron polar distribution in the muon rest frame. 
In the absence of the laser, the emission is completely isotropic. As the parameter $\Omega$ is increased, fringes build up in the laser wavevector direction and the rate is decreased overall. The higher $\Omega$ is made, the more numerous the fringes become. As $\Omega \to \infty$, the emission becomes isotropic again, but with half the rate. In the energy distribution of emitted electrons in the muon rest frame \figref{fig:EnergyPlot}, the suppression of the total rate can also be seen as $\Omega \to \infty$, although the appearance of fringes is much less pronounced, being clearest at the lowest energies.
After performing all momentum integrations, we find:
\bea 
\Wvac(\xi) = \mathcal{R}\left[\Omega(\xi)\right]\Wvac(0)
\eea 
where:
\bea 
\mathcal{R}[\Omega] &\approx& \begin{cases}
			1-\dfrac{5 \pi \Omega}{12} + \dfrac{\Omega^{2}}{18}\left(19 - 15 \gamma - 15 \log \Omega \right), & (\Omega \ll 1)\\[2ex]
            \dfrac{1}{2} - \dfrac{1}{\Omega^{2}} + \dfrac{20}{\Omega^{4}}  & (\Omega \gg 1)
		 \end{cases} \nn
\eea 
with $\gamma = -\int_{0}^{\infty}\mbox{e}^{-x}\ln x~dx \approx 0.577$ the Euler-Mascheroni constant. The decay rate in the absence of the laser is then the well-known result \cite{Griffiths:2008}, $\Wvac(0)=G^{2} m_{\mu}^{5}/192\pi^{3}$. (See \eqnref{eqn:Romega} in Supplementary B \cite{supp} for the explicit expression for $\mathcal{R}[\Omega]$.)
\begin{figure}[h!!]
	\includegraphics[width=6.5cm]{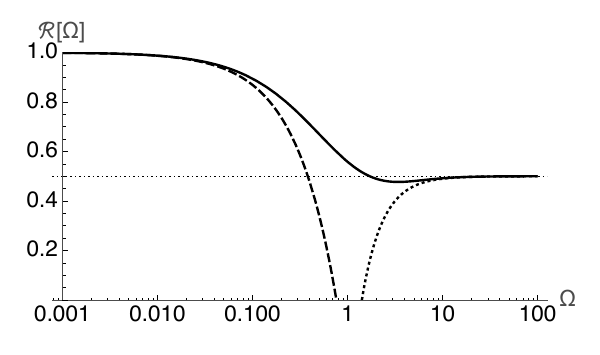}
\caption{A plot of the function $\mathcal{R}[\Omega]$ (solid line). The leading-order perturbative and asymptotic limits are indicated with dashed lines and the gridline is at $\mathcal{R}[\Omega] = 1/2$.} \label{fig:totRate}
\end{figure}
We note from \figref{fig:totRate} the effect of laser pulse interference on muon decay is a suppression of the rate, up to around half the vacuum value. Although the effect is quantum mechanical in nature, the parameter controlling pulse interference, $\Omega$, is entirely classical. Because the Volkov wavefunction is semiclassical exact, the nonlinear phase term from the muon and electron, $S_{a}(\Phi)$, can be understood as originating from the net change in the position of the muon and electron due to acceleration in the laser field. Explicitly, ${\Omega = p\cdot[x_{f}(\xi)-x_{f}(0)]}$ where $x_{f}(\xi)$ is the muon position after interacting with the laser pulse of intensity parameter $\xi$. By artificially `turning off' the muon-laser or electron-laser coupling by setting $\xi_{\mu}$ or $\xi_{e}$ to zero, we find that the dependence of the total rate most closely matches the electron-laser interaction (see Supplementary B \cite{supp}). This suggests that the decay of other heavy particles to electrons or positrons may be affected by the same mechanism. In very intense fields, or fields with a sufficiently long duration, classical radiation reaction may significantly modify a charge's trajectory in a plane wave background. Here, the radiation reaction parameter for the muon or electron is $\nu = (2/3)\alpha\eta\xi^{2}\Phi$.  If the field is very long, then eventually the probability, $\tsf{P}$, for the muon to decay, $\tsf{P}=\tsf{W}_{\tsf{vac}}T$, will increase and to maintain unitarity, higher orders in the weak-field coupling, $G$, such as loops, must be included. On the other hand, if the field is so intense that $\chi_{\mu} \sim O(1)$, then we would have to include the non-vacuum part of the probability that changes the decay kinematics, which we have neglected for reasons outlined in the introduction. However, we will see that the laser pulse does not need to be long or intense to affect muon decay, and since $\eta\Phi \ll 1$ this means $\nu \ll 1$ and therefore radiation reaction and channels with modified kinematics should not appreciably modify the result. We also note that the controlling parameter, $\Omega$, does not depend on the explicit pulse shape, but just as in the perturbative case, depends only on the integral of the square of the potential.

Recent experiments \cite{Calvin:2025huk} colliding laser wakefield accelerated electrons with solid targets created muons with energies $\approx 300\,\trm{MeV}$. If muons from such sources were collided with optical lasers, $\eta_{\mu}\sim O(10^{-8})$. Noting that $\av{g^{2}} \sim O(1)$ and the phase duration $\Phi = 2\pi N$ where $N \gg1$ is the number of laser cycles, writing $\xi_{\mu} = \delta\xi_{e}$, gives $\Omega \sim \pi N \delta^{2}\xi_{e}^{2}/\eta_{\mu}$. Clearly, one can reach $\Omega \gg 1$ even with weak fields, for which $\xi_{e} \ll 1$, implying that the suppression of vacuum muon decay could be observable using laser parameters available in today's facilities. If a broadband muon source, such as bremsstrahlung, is to be used in experiment, then it would be important to understand how the bandwidth of the muon wavepacket changes the laser's effect on vacuum decay. Using a toy model \cite{Aleksandrov:2020xop} of a Gaussian wavepacket of muons collidng head-on with the laser pulse, we find that the laser's effect is only lightly suppressed, at mainly in the perturbative regime of $\Omega \ll 1$. At moderate $\Omega$, for example $\Omega \gtrsim 2$, there is effectively no change even for a $100\%$ bandwidth muon wavepacket (see Supplementary B \cite{supp} for details). It would also be important to understand how focussing effects, in particular the localised nature of the muon-laser interaction point, influence the main result. Any experimental test would need to be able to select for muon decays originating from trajectories that crossed the interaction point whilst the pulse was at the focus. However, since the laser intensity can be much weaker than the those used in all-optical muon sources \cite{Schumaker:2018noe,Zhang:2024axy,Calvin:2025huk}, the laser beam can be defocussed to provide a larger and more persistent target for the muons.

Throughout, we have assumed that neither the muon nor the electron radiate when interacting with the weak laser pulse. The probability of Compton scattering can be estimated using literature expressions for the perturbative limit $\xi\ll1$ and $\eta \ll 1$ e.g. in a circularly-polarised background plane wave of finite extent \cite{Heinzl:2020ynb}. Then $\tsf{P}_{e\to e+\gamma} \approx 2\alpha \xi^{2}_{e}\Phi \av{g^{2}}/3$ and likewise for the muon with $e\to \mu$. Comparing probabilities, we find: 
\[
\Pvac \approx 1.5\times10^{-3}\mathcal{R}(\Omega)L[m];\quad \tsf{P}_{e\to e+\gamma} \approx  10^{-4}\bar{N}\bar{\xi}_{e}^{2}
\]
 where $L[m]$ is the distance in metres to the detector, $\Omega \sim 0.8 \bar{N}\bar{\xi}_{e}^{2}/\bar{\eta}_{\mu}$ and where parameters have been scaled by typical experimental values: $\bar{N} = N/10$, $\bar{\xi}_{e} = \xi_{e}/0.02$, $\bar{\eta}_{\mu} = \eta_{\mu}/10^{-8}$. For a high-power laser with wavelength $800\,\trm{nm}$, the central frequency is $1.55\,\trm{eV}$, in which case $\eta_{\mu}=10^{-8}$ would correspond to muons with a kinetic energy of $40\,\trm{MeV}$. An intensity parameter of $\xi_{e}=0.02$ would correspond to an intensity of $1.6 \times 10^{15}\,\trm{Wcm}^{-2}$ \cite{DiPiazza:2010mv} and $N=10$ laser cycles to a full-width-at-half-maximum pulse duration of $15\,\trm{fs}$ for a sine-squared pulse envelope. Therefore if $\xi_{e}$ is made small enough $\Omega \sim O(1)$ can still be achieved with electron Compton scattering much less probable than muon decay. 

\emph{Conclusion --} We have shown how muon decay can be significantly influenced by interaction with a laser pulse modelled as a plane wave of finite extent. Interactions with the electron and muon that involve zero \emph{net} momentum change (for example absorbing and emitting the same number of laser photons)  result in an interference with the standard vacuum decay channel. This interference of histories between the muon decaying with or without interacting with the laser pulse can modify the emitted electron momentum spectrum and increase the muon lifetime by up to a factor 2. Because it depends on the muon being able to decay outside the laser pulse, this mechanism circumvents well-known limitations for manipulating electroweak processes with strong electromagnetic fields as it requires neither a large strong-field parameter nor a large intensity; indeed the effect can be demonstrated in weak fields. This work focussed on muon decay but the same arguments clearly apply more generally to electroweak decays involving electromagnetic charges. All-optical experiments are using high-$Z$ solid targets to create charged pions and muons in the lab \cite{Schumaker:2018noe,Zhang:2024axy,Calvin:2025huk}; future work could involve using a weaker secondary laser to investigate the effect on the number of electroweak decays via the mechanism outlined here. This could be supported by improved modelling that includes the localised nature of the interaction point in collisions with focussed laser pulses.

\section*{Acknowledgments}
The authors thank Alexander Fedotov, Tom Heinzl and Anton Ilderton for helpful comments and acknowledge support from The Leverhulme Trust, Grant RPG-2023-285.

\vspace{0.5cm}

\onecolumngrid

\bibliography{current}



\section{Supplementary A: weak-field calculation}

\begin{figure}[h!!]
\includegraphics[width=5.0cm]{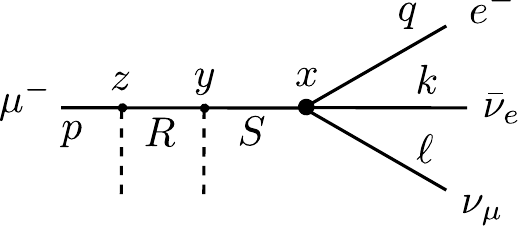}
\caption{Two interactions between the muon and the laser before decay into electron and neutrinos.} \label{fig:pertDiag1}
\end{figure}

In the main paper, we sum over six diagrams to obtain the amplitude for muon decay in a weak electromagnetic plane wave background to order $\xi^{2}$. By way of demonstration, we outline the derivation of $\tsf{T}_{20}$. We can write the partial amplitude as:
\bea 
\tsf{T}_{20} &=& \int d^{4}x \,d^{4}y\, d^{4}z\,\frac{d^{4}R}{(2\pi)^{4}}\,\frac{d^{4}S}{(2\pi)^{4}}~\mbox{e}^{ix\cdot (q+k+\ell)}\mathfrak{M}_{0}^{i} \frac{\slashed{R}+m}{R^{2}-m^{2}+i\epsilon} \mbox{e}^{-iR\cdot (x-y)} \slashed{a}(y)\frac{\slashed{S}+m}{S^{2}-m^{2}+i\epsilon}\slashed{a}(z)\mbox{e}^{-iS\cdot (y-z)}\mbox{e}^{-ip\cdot z}\frac{u_{p}}{\sqrt{2Vp^{0}}}\nn \\ \label{eqn:T20start}
\eea
where $a=eA$ and the `initial' part of the vacuum amplitude has been defined:
\bea 
\mathfrak{M}_{0}^{i} &=& \frac{G}{\sqrt{2}}\frac{1}{\sqrt{(2V)^{3}q^0k^0\ell^0}}\left[\bar{u}_{q}\gamma_{\rho}(1-\gamma_{5})v_{\ell}\right]~\bar{u}_{k}\gamma^{\rho}(1-\gamma_{5}) 
\eea
and is related to the total vacuum amplitude by $\mathfrak{M}_{00} = \mathfrak{M}_{0}^{i}u_{p}/\sqrt{2Vp^{0}}$.
 (A labelled diagram for this channel is given in  \figref{fig:pertDiag1}.)

Let us now specify to a plane wave, the profile of which we Fourier transform using:
\[
a^{\mu}(y_{+}) = m\xi\epsilon^{\mu}\, g(y_{+}); \qquad g(y_{+}) =  \int \frac{d\kappa^{+}}{2\pi}\,\tilde{g}(\kappa^{+})\mbox{e}^{i\kappa^{+}y_{+}},
\]
Upon substitution into \eqnref{eqn:T20start}, we find:
\bea 
\tsf{T}_{20}&=&8(m\xi)^{2}(2\pi)^{2}\,\int d^{4}R\,d^{4}S\,d\kappa^{+}\,d\kappa'^{+}~\delta^{\perp,+,-}\left(q+k+\ell-R\right)\delta^{\perp,+,-}\left(R-S-\kappa\right)\delta^{\perp,+,-}\left(S-p-\kappa'\right) \nn \\ 
&& \hspace{5cm}\times\, \mathfrak{M}_{0}^{i}\frac{\slashed{R}+m}{R^{2}-m^{2}+i\epsilon} \slashed{\epsilon}\frac{\slashed{S}+m}{S^{2}-m^{2}+i\epsilon} \slashed{\epsilon}\frac{u_{p}}{\sqrt{2Vp^{0}}}\tilde{g}(\kappa^{+})\tilde{g}(\kappa'^{+}).\nn \\
\eea
Integrating out the propagator momenta and Fourier-transforming back the profile, we find:
\bea 
\tsf{T}_{20}&=&2(m\xi)^{2}(2\pi)^{2}\delta^{\perp,-}(Q)\int dy_{+}\,dz_{+}\,d\kappa^{+}\,d\kappa'^{+}~\delta\left(Q^{+}-\kappa^{+}-\kappa'^{+}\right)\mbox{e}^{i\kappa^{+}y_{+}+i\kappa'^{+}z_{+}}\nn \\ 
&& \hspace{4cm} \times~ \mathfrak{M}_{0}^{i}\frac{\slashed{p}+\slashed{\kappa}+\slashed{\kappa}'+m}{2p_{+}(\kappa+\kappa')^{+}+i\epsilon} \slashed{\epsilon}\frac{\slashed{p}+\slashed{\kappa}'+m}{2p_{+}\kappa'^{+}+i\epsilon} \slashed{\epsilon}\frac{u_{p}}{\sqrt{2Vp^{0}}}g(y_{+})g(z_{+}).\nn \\ \label{eqn:T20rep2}
\eea
Employing the Sokhotski-Plemelj \cite{heitler60} theorem, we rewrite the propagator denominators and integrate out $\kappa'$ using the delta function:
\bea 
\tsf{T}_{20}&=&2(m\xi)^{2}\,(2\pi)^{2}\delta^{\perp,-}\left(Q\right)\int dy_{+}\,dz_{+}\,d\kappa^{+}~\mathfrak{M}_{0}^{i}\mbox{e}^{i\kappa^{+}y_{+}+i(Q^{+}-\kappa^{+})z_{+}}\nn \\ 
&& \times \frac{1}{p^{-}}\left\{\left(\slashed{p}+m\right)\left[-i\pi \delta(Q^{+}) + \widehat{P}\frac{1}{Q^{+}}\right]+ \gamma_{+}\right\}\slashed{\epsilon}\nn \\
&& \times \frac{1}{p^{-}}\left\{\left(\slashed{p}+m\right)\left[-i\pi \delta(Q^{+}-\kappa^{+}) + \widehat{P}\frac{1}{(Q-\kappa)^{+}}\right]+ \gamma_{+}\right\}\slashed{\epsilon}\frac{u_{p}}{\sqrt{2Vp^{0}}}\tilde{g}(y_{+})g(z_{+}).
\eea
We see contributions with vacuum kinematics $\propto\delta(Q^{+})$ and contributions from laser kinematics. Keeping just the vacuum kinematic term from the first propagator and integrating over the second, we acquire:
\bea 
\tsf{T}_{20}&=&-\frac{i}{2}\,(m\xi)^{2}\,(2\pi)^{4}\delta^{(4)}\left(Q\right)\frac{1}{p^{-}}\mathfrak{M}_{0}^{i}\left(\slashed{p}+m\right) \nn \\ 
&& \times \frac{1}{p^{-}}\left\{\slashed{\epsilon}\left(\slashed{p}+m\right)\slashed{\epsilon}\left[-i\int dy_{+}\,dz_{+}\theta\left(y^{-}-z^{-}\right)g(y_{+})g(z_{+})\right]+ \int dy_{+} \frac{\slashed{\epsilon}\slashed{\vkap}\slashed{\epsilon}}{\vkap^{+}}g^{2}(y_{+})\right\}\frac{u_{p}}{\sqrt{2Vp^{0}}}
\eea
(where we used: $2\delta^{\perp,-,+}(Q) = \delta^{(4)}(Q)$). Then using $\slashed{\epsilon}\slashed{\vkap}\slashed{\epsilon} = -\slashed{\epsilon}\slashed{\epsilon}\slashed{\vkap} = -\epsilon\cdot\epsilon \,\slashed{\vkap}$, reinstating $a(y)$ and $a(z)$ then and using $m u_{p} = p u_{p}$ and $\slashed{p}\slashed{a}+\slashed{a}\slashed{p} = 2p\cdot a$ and similar manipulation to show $(\slashed{p}+m)\slashed{\vkap}\,u_{p} = 2\vkap\cdot p\, u_{p}$, we find:
\bea 
\tsf{T}_{20}&=& \frac{1}{2}\tsf{T}_{00} \left\{\frac{1}{2}\left[i \int d\vphi~\frac{p\cdot a(\vphi)}{\vkap\cdot p}\right]^{2} + i  \int d\vphi \frac{a(\vphi)\cdot a(\vphi)}{2\,\vkap \cdot p}\right\} = \tsf{T}_{00} \frac{\xi^{2}}{2} \left[\frac{1}{2} \left(i\mathcal{I}_{1p}\right)^{2}+i\mathcal{I}_{2p}\right] \label{eqn:T20app3}
\eea
where $\tsf{T}_{00} = (2\pi)^{4}\delta^{(4)}\left(Q\right)\mathfrak{T}_{00}$. This should be compared to the Volkov case $\Tvac(\xi) = \Tvac(0) F(\xi)$ where:
\bea 
F(\xi) = \frac{1}{2}\left[1+\mbox{e}^{i \left(\xi \mathcal{I}_{1}+\xi^{2} \mathcal{I}_{2}\right)}\right] \approx 1 + \frac{i}{2} \xi \mathcal{I}_{1}  + \frac{\xi^{2}}{2}\left[-\frac{1}{2}\mathcal{I}_{1}^{2}+i\mathcal{I}_{2}\right]
\eea
where $\mathcal{I}_{1}=\mathcal{I}_{1q}-\mathcal{I}_{1p}$ and $\mathcal{I}_{2p}-\mathcal{I}_{2q}$.

\subsection{Decay Location}

To determine where the main contribution of muon decays occur, we can localise the muon decay in the above analysis, by making the replacement in \eqnref{eqn:T20start} of:
\[
\int_{-\infty}^{\infty} dx_{+} \to \int_{0}^{\tau_{+}} dx_{+}
\]
where $\tau_{+}$ is defined via the phase pulse duration $\Phi = \vkap^{+}\tau_{+}$, hence isolating the contribution from within the laser pulse. Then it follows that there is no momentum conservation in the $+$ lightfront component, due to the replacement in \eqnref{eqn:T20rep2} of:
\[
\delta^{+}\left(q+k+\ell-R\right) \to \frac{\tau_{+}}{\pi} \trm{sinc}\left[\tau_{+}\left(q^{+}+k^{+}+\ell^{+}-R^{+}\right)\right].
\]
The rest of the integrals can be performed as before, such that we can write the localised version of \eqnref{eqn:T20app3} from `inside' the pulse compared to `outside' the pulse:
\bea
\tsf{T}^{\tsf{outside}}_{20} &=& 2(2\pi)^{4}\delta^{\perp,-,+}\left(Q\right)\mathfrak{T}_{00} \frac{\xi^{2}}{2} \left[\frac{1}{2} \left(i\mathcal{I}_{1p}\right)^{2}+i\mathcal{I}_{2p}\right]\nn \\
\tsf{T}^{\tsf{inside}}_{20} &=& 2(2\pi)^{4}\delta^{\perp,-}\left(Q\right)\frac{\tau_{+}}{\pi}\trm{sinc}(\tau_{+}Q^{+})\mathfrak{T}_{00} \frac{\xi^{2}}{2} \left[\frac{1}{2} \left(i\mathcal{I}_{1p}\right)^{2}+i\mathcal{I}_{2p}\right].\nn 
\eea
We see that at the probability level when we mod-square the amplitude that contribution to the probability from the pulses scales as:
\[
\Big|\tsf{T}^{\tsf{outside}}_{20}\Big|^{2} \sim VT; \qquad \Big|\tsf{T}^{\tsf{inside}}_{20}\Big|^{2} \sim VT \frac{\tau_{+}}{T_{+}} \frac{\tau_{+}}{\pi^{2}}\trm{sinc}^{2}(\tau_{+}Q^{+}).
\]
where $T_{+}$ is the normalisation lightfront time defined through the relation:
\[
\Big|2\delta^{\perp,-,+}\left(Q\right)\Big|^{2} =  \frac{4T_{+}T_{-}A}{(2\pi)^{4}} \delta^{\perp,-,+}\left(Q\right) \equiv \frac{VT}{(2\pi)^{4}} \delta^{\perp,-,+}\left(Q\right)
\]
(where $A$ is the normalisation area). We see that, even in the case that $\trm{sinc}^{2}$ oscillates slowly as $Q^{+}$ is integrated over, there is still the prefactor of $\tau_{+}/T_{+}$ which is approximately the ratio of the pulse duration to the time of flight to the detector, which we take throughout to be very small. Therefore, we conclude the contribution from inside the pulse, whilst not zero, is much smaller.

\section{Supplementary B: Strong-field calculation} 
First, recall from the main paper that the transition matrix amplitude can be written as:
\[
\Tfi = \int d^{4}x \,\mbox{e}^{i Q\cdot x} \,\Mfi
\]
where the invariant amplitude:
\[
\Mfi = \frac{G}{\sqrt{2}}\,J_{q,\ell,\sigma}J_{p,k}^{\sigma};\qquad 
J_{q,\ell,\sigma} = \frac{1}{\sqrt{(2V)^{2}q^{0}\ell^{0}}}\bar{u}_{q}\gamma_{\sigma}(1-\gamma_{5})v_{\ell}; \qquad J_{p,k}^{\sigma} = \frac{1}{\sqrt{(2V)^{2}p^{0}k^{0}}}\bar{u}_{k}\gamma^{\sigma}(1-\gamma_{5})u_{p}.
\]
Employing Volkov states to describe the charge-field interaction, the amplitude now becomes `dressed' in the background field, i.e. $J_{q,\ell,\sigma}= J_{q,\ell,\sigma}(a)$ and $J_{p,k}^{\sigma}=J_{p,k}^{\sigma}(a)$ due to the replacement:
\bea
u_{p}\mbox{e}^{-i p\cdot x} \to \left[1+\frac{\slashed{\vkap}\slashed{a}(\phi)}{2\vkap\cdot p}\right]u_{p}\mbox{e}^{-ip\cdot x + iS_{a,p}(\phi)}; \qquad S_{a,p}(\phi) = \int^{\phi}_{-\infty} \left[\frac{-p \cdot a(\vphi)}{\vkap \cdot p} + \frac{a(\vphi)\cdot a(\vphi)}{2\vkap \cdot p}\right] d\vphi \label{eqn:muonWF1}
\eea
and likewise for $\bar{u}_{q}$. This leads to a transition matrix amplitude of the form:
\[
\Tfi = \int d^{4}x \,\mbox{e}^{i Q\cdot x + iS_{a}(\phi)} \,\Mfi[a]
\]
where $Q=q+\ell + k - p$ is the total momentum change and $S_{a}(\phi)=S_{a,p}(\phi)-S_{a,q}(\phi)$. We are interested in the vacuum decay amplitude, and so Taylor expand $\Mfi[a]$ in $a$ taking the zero order only. (As explained in the main text, we discard the other terms involving the laser pulse due to the vanishing contribution in the regime of interest.)  Calling the remaining amplitude $\Tfi^{(0)}$ we see:
\bea 
\Tfi^{(0)} &=& 2(2\pi)^{3}\delta^{\perp,-}(Q)\Mfi[0]\int_{-\infty}^{\infty} dx_{+} \,\mbox{e}^{i Q^{+}x_{+} + iS_{a}(\phi)}.
\eea
The integral is then split into three parts: before the pulse, during the pulse and after the pulse. Rearranging, we have:
\bea 
\int_{-\infty}^{\infty} dx_{+} \,\mbox{e}^{i Q^{+}x_{+} + iS_{a}(\phi)} &=& \int_{-\infty}^{0} dx_{+} \,\mbox{e}^{i Q^{+}x_{+}} + \int_{0}^{\Phi} dx_{+} \,\mbox{e}^{i Q^{+}x_{+} + iS_{a}(\phi)} + \int_{\Phi}^{\infty} dx_{+} \,\mbox{e}^{i Q^{+}x_{+} + iS_{a}(\Phi)}\nn \\
&=& 2\pi \delta\left(Q^{+}\right) + \left[\mbox{e}^{iS_{a}(\Phi)}-1\right]\lim_{\eps\to0}\int_{0}^{\infty} dx_{+} \,\mbox{e}^{i (Q^{+}+i\eps)x_{+}} + \int_{0}^{\Phi} dx_{+} \,\mbox{e}^{i Q^{+}x_{+}}\left[\mbox{e}^{iS_{a}(\phi)}-\mbox{e}^{iS_{a}(\Phi)}\right]. \nn  \eea 
Performing the middle integral and using the Sokhotski-Plemelj theorem \cite{heitler60}, the expression can be recast in the form:
\bea
\int_{-\infty}^{\infty} dx_{+} \,\mbox{e}^{i Q^{+}x_{+} + iS_{a}(\phi)} &=& 
 2\pi F(a)\delta\left(Q^{+}\right) + \left[\mbox{e}^{iS_{a}(\Phi)}-1\right]i\hat{P}\frac{1}{Q^{+}}+ \int_{0}^{\Phi} dx_{+} \,\mbox{e}^{i Q^{+}x_{+}}\left[\mbox{e}^{iS_{a}(\phi)}-\mbox{e}^{iS_{a}(\Phi)}\right] \nn
\eea
\bea
F(a) = \frac{1+\mbox{e}^{iS_{a}(\Phi)}}{2}; \qquad |F(a)|^{2} = \frac{1+\cos \left[S_{a}(\Phi)\right]}{2}
\eea
and $F(0)=1$. Altogether therefore, we have $\Tfi = (2\pi)^{4}\delta^{(4)}(Q)\Mfi^{(0)} F(a) + 2(2\pi)^{3}\delta^{\perp,-}(Q)\left(\cdots\right)$. When the probability is formed $\Tfi$ is mod-squared and contains interference between the vacuum and `laser' channels. However, because of the different kinematics involved, and the integration over the pulse phase in the laser channel, contributions to the total probability that are not entirely from the vacuum channel, scale with the pulse duration $\tau = \Phi/\vkap$ compared to the muon time of flight to the detector $T$. Since $T\gg \tau$, we drop all other contributions that are not the purely vacuum channel i.e. assume $\Tfi = \Tvac(a) = \Tvac(0) F(a)$ with $\Tvac(0)= (2\pi)^{4}\delta^{(4)}(Q)\mathfrak{M}_{00}$. The rate of muon decay $\Wvac = \Pvac / T$ (where $\Pvac$ is the muon decay probability) is then:
\bea 
\Wvac = \frac{V^{3}}{T}\int \frac{d^{3}q\,d^{3}\ell\,d^{3}k}{(2\pi)^{9}}\,\frac{1}{2}(2\pi)^{8}|\mathfrak{M}_{00}|^2\left\{1 + \cos\left[S_{a}(\Phi;q^{-})\right]\right\}
\eea 
where we write $S_{a}(\Phi)$ as $S_{a}(\Phi;q^{-})$ here and in the following to emphasise the momentum dependency of the phase in the integral. If $|\mathfrak{M}_{00}|^{2}$ corresponds to the \emph{unpolarised} probability, then multiplying by a factor $1/2$ to average over the initial spin of the muon, we find:
\bea 
\frac{1}{2}\sum_{\sigma_{p},\sigma_{q}}\tsf{tr}~ J_{q,\ell,\mu}J^{\mu}_{p,k} = 2^{7}\,k\cdot q\,\ell \cdot p\,, \qquad \trm{implying} \qquad |\mathfrak{M}_{00}|^{2} = \frac{2^{6}G^{2}}{(2V)^{4}p^{0}q^{0}\ell^{0}k^{0}}~k\cdot q\,\ell \cdot p.
\eea
Squaring the delta function: $\left[\delta^{4}\left(q+\ell+k-p\right)\right]^{2} = [VT/(2\pi)^{4}]~\delta^{4}\left(q+\ell+k-p\right)$, we then have:
\bea 
\Wvac = \frac{2G^{2}}{(2\pi)^{5}p^{0}}\int \frac{d^{3}q\,d^{3}\ell\,d^{3}k}{q^{0}\ell^{0}k^{0}}~k\cdot q~\ell\cdot p~\left\{1+\cos[S_{a}(\Phi;q^{-})]\right\}\delta^{4}(q+\ell+k-p).
\eea
At this point, we mainly follow the derivations in Griffiths \cite{Griffiths:2008}. Since the field-dependent term only depends on the $q^{-}$ integration variable, we leave this integration to last. First, we 
use the delta-function to integrate out $d^{3}\mbf{k}$:
\bea 
\Wvac = \frac{2G^{2}}{(2\pi)^{5}p^{0}}\int \frac{d^{3}q\,d^{3}\ell}{q^{0}\ell^{0}k^{0}_{\ast}}~k_{\ast}\cdot q~\ell\cdot p~\left\{1+\cos[S_{a}(\Phi;q^{-})]\right\}\delta\left(q^{0}+\ell^{0}+k^{0}_{\ast}-p^{0}\right)
\eea
where we use the asterisk notation to denote quantities that have already been integrated out and can be expressed in the remaining integration variables, here for example $\tbf{k}_{\ast} = \mbf{p}_{\ast}-\mbf{q}_{\ast}-\pmb{\ell}_{\ast}$ with the final component of momentum $k^{0}_{\ast}=\sqrt{m_{k}^{2}+|\mbf{p}-\mbf{q}-\pmb{\ell}|^2}$ fixed by the on-shell condition. Switching to the muon rest frame so that $p=(m_{p},\pmb{0})$, we use spherical polars for $d^{3}\pmb{\ell} = |\pmb{\ell}|^2\,\sin\theta_{\ell}\,d|\pmb{\ell}|\,d\phi_{\ell}\,d\theta_{\ell}$ and write:
\[
k^{0}_{\ast}= \sqrt{m_{k}^{2}+|\mbf{q}|^{2}+|\pmb{\ell}|^2+2|\mbf{q}||\pmb{\ell}|\cos\theta_{\ell}},
\]
i.e. choosing the $\pmb{\ell}$ axes so the polar angle $\theta_{\ell}$ is equal to the angle between $\mbf{q}$ and $\pmb{\ell}$. Defining the integration variable:
\[
x = k^{0}_{\ast}; \qquad dx = -\frac{|\mbf{q}||\pmb{\ell}|\sin \theta_{\ell}}{x}\,d\theta_{\ell},
\]
and substituting, gives:
\bea 
\Wvac = \frac{2G^{2}}{(2\pi)^{5}p^{0}}\int \frac{d^{3}q\,dx\,d\phi_{\ell}d|\pmb{\ell}|\,|\pmb{\ell}|}{q^{0}\ell^{0}|\mbf{q}|}~k\cdot q~\ell\cdot p~\left(1+\cos[S_{a}(\Phi;q^{-})]\right)\delta\left(q^{0}+\ell^{0}-m_{p}+x\right).
\eea
Since $-1<\cos\theta_{\ell} <1$, there is a condition on other integration variables
\bea
\cos\theta_{\ell}&=&\frac{\left[m_{p}-(q^0+\ell^0)\right]^2-(m_{k}^{2}+|\mbf{q}|^{2}+|\pmb{\ell}|^2)}{2|\mbf{q}||\pmb{\ell}|} \nn \\
&=&\frac{1}{2|\mbf{q}||\pmb{\ell}|}\left(|m_{p}-(q^0+\ell^0)|-\sqrt{m_{k}^{2}+|\mbf{q}|^{2}+|\pmb{\ell}|^2}\right)\left(|m_{p}-(q^0+\ell^0)|+\sqrt{m_{k}^{2}+|\mbf{q}|^{2}+|\pmb{\ell}|^2}\right).\nn \\
\eea
For the delta-function to evaluate to a non-zero value and considering the maximum and minimum values that $\cos\theta_{\ell}$ can take, we see: 
\[
x_{\trm{min}} < m_{p}-(q^0+\ell^0) < x_{\trm{max}}
\]
\[
\sqrt{m_{k}^{2}+(|\mbf{q}|-|\pmb{\ell}|)^{2}}< m_{p}-(q^0+\ell^0)<\sqrt{m_{k}^{2}+(|\mbf{q}|+|\pmb{\ell}|)^{2}}.
\]
Adding $q^{0}+\ell^{0}$ and dividing by two:
\[
\frac{\sqrt{m_{k}^{2}+(|\mbf{q}|-|\pmb{\ell}|)^{2}}+q^0+\ell^0}{2}< \frac{m_{p}}{2}<\frac{\sqrt{m_{k}^{2}+(|\mbf{q}|+|\pmb{\ell}|)^{2}}+q^0+\ell^0}{2}.
\]
If we set the neutrino masses to be zero at this point, we find the condition:
\[
\frac{m_{p}-q^{0}-|\mbf{q}|}{2} < \ell^{0} < \frac{m_{p}-q^{0}+|\mbf{q}|}{2}\,.
\]
These inequalities allow us to place bounds on the $\ell^{0}$ integral. We use:
\[
y = (m_{\ell}^{2}+|\pmb{\ell}|^2)^{1/2} = \ell^{0}; \qquad dy = \frac{|\pmb{\ell}|\,d|\pmb{\ell}|}{y}\, .
\]
Then performing the $dx$ and $\phi_{\ell}$ integrals, we have:
\bea 
\Wvac = \frac{2G^{2}}{(2\pi)^{4}}\int \frac{d^{3}q}{q^{0}|\mbf{q}|}\int^{(m_{p}-q^{0}+|\mbf{q}|)/2}_{(m_{p}-q^{0}-|\mbf{q}|)/2}dy\,y ~k\cdot q~\left(1+\cos[S_{a}(\Phi;q^{-})]\right)
\eea
where we used $\ell\cdot p = \ell^{0}p^{0}=yp^{0}=ym_{p}$ and cancelled a factor of $1/l^{0}$ inside and $1/p^{0}$ outside the integral. The remaining dot-product is:
\bea
k\cdot q &=& k^{0}q^{0}-\mbf{k}\cdot\mbf{q} = q^{0}x+\mbf{q}\cdot(\mbf{q}+\pmb{\ell}) = q^{0}x+|\mbf{q}|^2 +\mbf{q}\cdot\pmb{\ell},
\eea
and using:
\bea
x^{2} - m_{k}^{2} &=& \mbf{q}^{2} + \pmb{\ell}^{2} + 2\mbf{q}\cdot \pmb{\ell} \nn \\
\left[m_{p}-\left(q^{0}+y\right)\right]^{2}-m_{k}^{2} &=& \mbf{q}^{2} +y^{2}-m_{\ell}^{2} + 2\mbf{q}\cdot \pmb{\ell} \nn \\
\left[m_{p}-q^{0}\right]^{2}-m_{k}^{2} -2y(m_{p}-q^{0}) -\mbf{q}^{2}+m_{\ell}^{2}&=& 2\,\mbf{q}\cdot \pmb{\ell}
\eea
the remaining dot-product can be written:
\bea 
k\cdot q &=& \mbf{q}^{2} + q^{0}\left[m_{p}-(q^{0}+y)\right] + \frac{\left[m_{p}-q^{0}\right]^{2}-m_{k}^{2} -2y(m_{p}-q^{0}) -\mbf{q}^{2}+m_{\ell}^{2}}{2} 
= -m_{p}y + \frac{m_{p}^{2}-m_{q}^{2}}{2}
\eea
where in the final line, we have used the fact that the neutrino masses have been set equal to zero. Performing the $y$ integral gives:
\bea 
\Wvac = \frac{2G^{2}}{(2\pi)^{4}}\int \frac{d^{3}q}{q^{0}|\mbf{q}|}\,\left[-\frac{m_{p}|\mbf{q}|(|\mbf{q}|^{2}+3(m_{p}-q^{0})^{2})}{12}+\frac{(m_{p}^{2}-m_{q}^{2})|\mbf{q}|(m_{p}-q^{0})}{4}\right]~\left(1+\cos[S_{a}(\Phi;q^{-})]\right).
\eea
The electromagnetic field dependence is in the cosine phase. Let us define:
\bea 
\cos\left[S_{a}(\Phi;q^{-})\right] &=& \cos \left\{\left[\frac{1}{2\vkap \cdot p} - \frac{1}{2\vkap \cdot q}\right]C_{\Phi}\right\}; \qquad C_{\Phi} = \int^{\Phi} a \cdot a ~ d\phi \,. \nn
\eea
We can write $d^{3}\mbf{q}$ in polar co-ordinates, now defining the $\mbf{q}$ polar co-ordinate, $\theta_{q}$, via the dot product in the field-dependent term:
\[
\vkap \cdot q = \vkap^{0}\left(q^{0}-|\mbf{q}|\cos\theta_{q}\right). 
\]
We use $d^{3}\mbf{q} = -|\mbf{q}|^{2} \,d|\mbf{q}|\,d\left(\cos\theta_{q}\right)d\phi_{q}$ and substitute the integration in the radial direction by defining:
\[
z = \left(m_{q}^2+|\mbf{q}|^2\right)^{1/2} = q^{0}; \qquad z\, dz = |\mbf{q}|\,d|\mbf{q}|
\]
giving $d^{3}\mbf{q} = - q^{0}|\mbf{q}|\, dz\, dX\,d\phi_{q}$ where $X = \cos\theta_{q}$. Then we have:
\bea 
\Wvac &=& \frac{2G^{2}}{(2\pi)^{4}}\int_{m_{q}}^{m_{p}/2} dz \int_{-1}^{1} dX\,\int d\phi_{q}~\left(z^{2}-m_{q}^{2}\right)^{1/2}\left[\frac{-m_{p}(z^{2}-m_{q}^{2}+3(m_{p}-z)^{2})}{12}+\frac{(m_{p}^{2}-m_{q}^{2})(m_{p}-z)}{4}\right] \nn \\
&& \!\!\!\!\times \left(1+\cos \left[\frac{C_{\Phi}}{2\vkap \cdot p}\right]\cos\left[ \frac{C_{\Phi}}{2\vkap^{0}\left(z-X\sqrt{z^{2}-m_{q}^{2}}\right)}\right]+\sin\left[\frac{C_{\Phi}}{2\vkap \cdot p}\right]\sin\left[\frac{C_{\Phi}}{2\vkap^{0}\left(z-X\sqrt{z^{2}-m_{q}^{2}}\right)}\right] \right) \label{eqn:WvacAppB1}
\eea
Usually, the electron mass does not significantly influence the muon decay rate and is set to zero. We investigate this point in an external field by replacing $m_{q} \to \delta m_{p}$ and expand in $\delta$. First, in the zero-field limit, we find:
\bea 
\lim_{\xi \to 0}\Wvac &=& \frac{G^{2}T m_{p}^{5}}{192\pi^{3}}f(\delta); \qquad f(\delta) = 24\delta^{4}\ln\left(\frac{2\delta}{1-\sqrt{1-4\delta^{2}}}\right) + \left[1-8\delta^{2}(1+\delta^{2})\right]\sqrt{1-4\delta^{2}}
\eea
with $f(\delta) \approx 1 - 10\,\delta^{2}$ for $\delta \ll 1$. For the electron, $\delta \approx 1/207$ and we find $1-f(1/207)=2.3 \times 10^{-4}$. For the field-dependent terms in the integrand, it is possible to perform the $z$ integral analytically. Taylor expanding in $\delta$ coefficients of trigonometric functions, we find that all corrections $O(\delta)$ are multiplied by fast-oscillating cosine and sine functions with arguments $\Omega/[\delta(1-X)]$ and only at $O(\delta^{2})$ do terms without fast oscillations occur (as in the $\xi \to 0$ case). Therefore we set the electron mass equal to zero in what follows.

In the muon rest frame, $2\vkap \cdot p = 2\vkap^{0}m_{p}$, so defining: $Y = C_{\Phi}/2\vkap^{0}m_{p}$ and rescaling $z=m_{p}Z$, we have:
\bea 
\Wvac &=& \frac{G^{2}}{(2\pi)^{3}}\frac{m_{p}^{5}}{6}\int_{0}^{1/2} dZ \int_{-1}^{1} dX\,Z^{2}(3-4Z)\left(1+\cos(Y)\,\cos\left[ \frac{Y}{Z(1-X)}\right]+\sin(Y)\,\sin\left[\frac{Y}{Z(1-X)}\right] \right) 
\eea
The remaining integrals can be performed analytically to give:
\bea 
\Wvac(\xi) = \mathcal{R}\left[\Omega(\xi)\right]\Wvac(0);  \qquad \Wvac(0) = \frac{G^{2}m_{p}^{5}}{192\pi^{3}} \label{eqn:WvacAppB1}
\eea
where:
\bea
\mathcal{R}\left[\Omega\right] &=& 1+\frac{\Omega}{72}\left\{16\Omega-3\pi(10+3\Omega^{2})\cos \Omega+\pi \Omega^{3}\sin \Omega - 2\,\trm{Ci}(\Omega) \left[\Omega^{3}\cos \Omega +3(10+3\Omega^{2})\sin \Omega\right] \right. \nn \\
&& \left. +2\, \trm{Si}(\Omega)\left[3(10+3\Omega^{2})\cos \Omega - \Omega^{3}\sin \Omega\right] \right\} \label{eqn:Romega}
\eea
and the result, derived in the muon rest frame, has been written in  frame-independent form by replacing $Y\to \Omega$ with $\Omega = C_{\Phi}/2 \vkap\cdot p$. The expression in \eqnref{eqn:WvacAppB1} is the same as in the main paper (when the replacement $m_{p} \to m_{\mu}$ is made).

\subsection{Charge-Laser Interaction}
\begin{figure}[h!!]
\centering
\includegraphics[width=7.5cm]{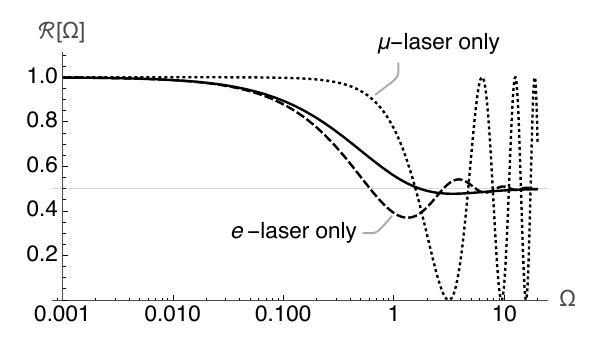}
\caption{How the rate of the vacuum channel of muon decay depends on $\Omega$ when only the electron-laser interaction is included (dashed line) or when only the muon-laser interaction is included (dotted line) compared with full interaction with both electron and muon (solid line).} \label{Fig:whichPlot1}
\end{figure}
We can investigate whether the muon-laser or electron-laser interaction is more important in the vacuum muon decay channel by artificially `turning off' the muon-laser or electron-laser interaction by setting $\xi_{\mu}\to 0$ or $\xi_{e}\to 0$. This corresponds to setting $C_{\Phi}/2\vkap \cdot p \to 0$ or  $C_{\Phi}/2\vkap \cdot q \to 0$ in \eqnref{eqn:WvacAppB1} and performing the integration. For the muon-laser-only interaction the result is straightforward and $\mathcal{R}[\Omega] \to (1+\cos\Omega)/2$. The electron-laser-only interactions requires further integration. From \figref{Fig:whichPlot1}, we see that the full dependency of $\mathcal{R}$ is well-described by just including the electron-laser interaction only.

\subsection{Muon wavepacket}
In this section we investigate the effect on the laser-muon decay mechanism if the muon is localised in a wavepacket. We multiply the muon wavefunction in \eqnref{eqn:muonWF1} with the envelope $\rho(p^{-})$ where:
\[
\rho(p^{-}) = \frac{1}{\sqrt{\Delta p^{-} \sqrt{\pi}}}\mbox{e}^{-\frac{1}{2}\left(\frac{p^{-}}{\Delta p^{-}}\right)^{2}};\qquad \int_{-\infty}^{\infty} |\rho(p^{-})|^{2} dp^{-} = 1
\]
adapting the approach in \cite{Aleksandrov:2020xop} for a muon wavepacket colliding head-on with the laser pulse, which has central wavevector $\vkap=\vkap^{+}$. Following the derivation for the plane-wave case, the major difference now is in the conserved momenta at amplitude level:
\bea
\Tfi &=& \int dp^{-} \rho(p^{-}) (2\pi)^{4}\delta^{(4)}(Q)\Mfi^{(0)} F(a) = 2(2\pi)^{3}\rho\left(p_{\trm{out}}^{-}\right)\delta^{\perp,+}(Q)\left[\Mfi^{(0)} F(a)\right]_{p^{-}=p^{-}_{\trm{out}}}
\eea
where $p^{-}_{\trm{out}} = q^{-}+k^{-}+\ell^{-}$ replaces the component $p^{-}$ in all parts of the amplitude. The derivation used for the plane wave muon cannot be straightforwardly adapted to this case because the spread of momenta in the muon wavepacket means there is no single `rest frame' for the muon. Because the muon wavepacket depends in a nonlinear way on the momenta of all emitted particles, the integrand can no longer be written just in terms of the radii and polar angles in momentum spherical co-ordinates.  Instead, we will use lightfront co-ordinates and then a numerical approach. Writing:
\bea 
\left[\delta^{\perp,+}\left(q+\ell+k-p\right)\right]^{2} 
&=& \frac{VT}{(2\pi)^{3}L_{-}}\delta^{\perp,+}\left(q+\ell+k-p\right),
\eea
we see the volumetric factors are different compared to the plane-wave muon case. However, the factors will be the same whether the laser is switched on or not, and so we can still calculate the ratio $\mathcal{R}$ and compare the effect of the laser. 
\begin{figure}[h!!]
\centering
\includegraphics[width=6cm]{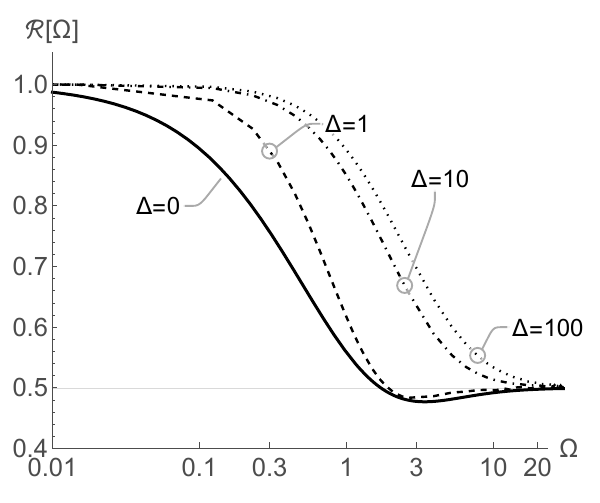}
\caption{How the influence of the laser on the vacuum muon decay channel changes as a function of the bandwidth $\Delta$ of the muon wavepacket, as a ratio of the muon's initial lightfront momentum.} \label{fig:muonPacket}
\end{figure}
Finally, we arrive at:
\bea 
\tsf{P} &=& \frac{2\pi}{L_{-}} \frac{T}{(2\pi)^{5}}\frac{2G^{2}}{p^{0}}\int \frac{d^{3}q\,d^{3}\ell}{q^{0}\ell^{0}k^{+}_{\ast}}~k_{\ast}\cdot q~\ell\cdot p~\left\{1+\cos[f_{a}(q^{-};\Phi)]\right\}\,\rho\left(p^{-}_{\trm{out}}\right)
\eea
where $k_{\ast}^{+,\perp}=p^{+,\perp}-q^{+,\perp}-\ell^{+,\perp}$ and $k_{\ast}^{-}=\mbf{k}^{\perp}_{\ast}\cdot\mbf{k}^{\perp}_{\ast}/k_{\ast}^{+}$, which we evaluate numerically. The integral over lightfront momenta is now over a product of interference factor and muon momentum wavepacket. Defining $\Delta = \Delta p^{-}/p^{-}$, we find the dependency in \figref{fig:muonPacket}. The plane-wave result corresponds to the limit $\Delta \to 0$. As the bandwidth is increased, larger values of $\Omega$ are required to significantly modify muon decay. The biggest effect is at small $\Omega$, but even at a bandwidth of $100\%$, i.e. $\Delta = 1$, by $\Omega \approx 2$, the dependency is the same for plane waves.

\end{document}